\documentclass[%
reprint,
superscriptaddress,
 amsmath,amssymb,
 aps,
onecolumn
]{revtex4-2}
\usepackage{titlesec}
\titleformat{\section}{\normalfont\bfseries}{\thesection}{1em}{}
\usepackage{graphicx}
\usepackage{dcolumn}
\usepackage{bm}
\usepackage{esvect}
\usepackage{hyperref}
\usepackage{booktabs}
\usepackage{soul}
\usepackage{xcolor}
\usepackage{diagbox}
\usepackage{makecell}
\usepackage{array}
\usepackage{upgreek}
\usepackage{esvect}
\usepackage{caption}
\usepackage{ulem}
\usepackage{ragged2e}

\usepackage[defaultcolor=blue]{changes}
\usepackage[T1]{fontenc}
\begin{document}

\setlength{\parindent}{0pt}
\preprint{APS/123-QED}

\title{High-fidelity geometric quantum gates exceeding 99.9$\%$ in germanium quantum dots}

\author{Yu-Chen Zhou}
\affiliation{Key Laboratory of Quantum Information, University of Science and Technology of China, Hefei, Anhui 230026, China}
\affiliation{CAS Center for Excellence in Quantum Information and Quantum Physics, University of Science and Technology of China, Hefei, Anhui 230026, China}
\author{Rong-Long Ma}
\affiliation{Key Laboratory of Quantum Information, University of Science and Technology of China, Hefei, Anhui 230026, China}
\affiliation{CAS Center for Excellence in Quantum Information and Quantum Physics, University of Science and Technology of China, Hefei, Anhui 230026, China}
\author{Zhenzhen Kong}
\affiliation{Integrated Circuit Advanced Process R$\&$D Center, Institute of Microelectronics, Chinese Academy of Sciences, Beijing 100029, China}
\affiliation{Beijing Superstring Academy of Memory Technology, Beijing 100176, China}
\author{Ao-Ran Li}
\affiliation{Key Laboratory of Quantum Information, University of Science and Technology of China, Hefei, Anhui 230026, China}
\affiliation{CAS Center for Excellence in Quantum Information and Quantum Physics, University of Science and Technology of China, Hefei, Anhui 230026, China}
\author{Chengxian Zhang}
\affiliation{School of Physical Science and Technology, Guangxi University, Nanning 530004, China}
\author{Xin Zhang}
\affiliation{QuTech and Kavli Institute of Nanoscience, Delft University of Technology, Lorentzweg 1, 2628 CJ Delft, The Netherlands}
\author{Yang Liu}
\affiliation{Key Laboratory of Quantum Information, University of Science and Technology of China, Hefei, Anhui 230026, China}
\affiliation{CAS Center for Excellence in Quantum Information and Quantum Physics, University of Science and Technology of China, Hefei, Anhui 230026, China}
\author{Hao-Tian Jiang}
\affiliation{Key Laboratory of Quantum Information, University of Science and Technology of China, Hefei, Anhui 230026, China}
\affiliation{CAS Center for Excellence in Quantum Information and Quantum Physics, University of Science and Technology of China, Hefei, Anhui 230026, China}
\author{Zhi-Tao Wu}
\affiliation{Key Laboratory of Quantum Information, University of Science and Technology of China, Hefei, Anhui 230026, China}
\affiliation{CAS Center for Excellence in Quantum Information and Quantum Physics, University of Science and Technology of China, Hefei, Anhui 230026, China}
\author{Gui-Lei Wang}
\email{guilei.wang@bjsamt.org.cn}
\affiliation{Integrated Circuit Advanced Process R$\&$D Center, Institute of Microelectronics, Chinese Academy of Sciences, Beijing 100029, China}
\affiliation{Beijing Superstring Academy of Memory Technology, Beijing 100176, China}
\affiliation{Hefei National Laboratory, Hefei 230088, China}
\author{Gang Cao}
\affiliation{Key Laboratory of Quantum Information, University of Science and Technology of China, Hefei, Anhui 230026, China}%
\affiliation{CAS Center for Excellence in Quantum Information and Quantum Physics, University of Science and Technology of China, Hefei, Anhui 230026, China}
\affiliation{Hefei National Laboratory, Hefei 230088, China}
\author{Guang-Can Guo}
\affiliation{Key Laboratory of Quantum Information, University of Science and Technology of China, Hefei, Anhui 230026, China}%
\affiliation{CAS Center for Excellence in Quantum Information and Quantum Physics, University of Science and Technology of China, Hefei, Anhui 230026, China}
\affiliation{Hefei National Laboratory, Hefei 230088, China}
\author{Hai-Ou Li}
\email{haiouli@ustc.edu.cn}
\affiliation{Key Laboratory of Quantum Information, University of Science and Technology of China, Hefei, Anhui 230026, China}%
\affiliation{CAS Center for Excellence in Quantum Information and Quantum Physics, University of Science and Technology of China, Hefei, Anhui 230026, China}
\affiliation{Hefei National Laboratory, Hefei 230088, China}
\author{Guo-Ping Guo}
\email{gpguo@ustc.edu.cn}
\affiliation{Key Laboratory of Quantum Information, University of Science and Technology of China, Hefei, Anhui 230026, China}%
\affiliation{CAS Center for Excellence in Quantum Information and Quantum Physics, University of Science and Technology of China, Hefei, Anhui 230026, China}
\affiliation{Hefei National Laboratory, Hefei 230088, China}
\affiliation{Origin Quantum Computing Company Limited, Hefei, Anhui 230088, China}






\date{\today}
\begin{abstract}
\noindent
Achieving high-fidelity and robust qubit manipulations is a crucial requirement for realizing fault-tolerant quantum computation. Here, we demonstrate a single-hole spin qubit in a germanium quantum dot and characterize its control fidelity using gate set tomography. The maximum control fidelities reach 97.48$\%$, 99.81$\%$, 99.88$\%$ for the $I$, $X$/2 and $Y$/2 gate, respectively. These results reveal that off-resonance noise during consecutive $I$ gates in gate set tomography sequences severely limits qubit performance. Therefore, we introduce geometric quantum computation to realize noise-resilient qubit manipulation. The geometric gate control fidelities remain above 99$\%$ across a wide range of Rabi frequencies. The maximum fidelity surpasses 99.9$\%$. Furthermore, the fidelities of geometric $X$/2 and $Y$/2 ($I$) gates exceed 99$\%$ even when detuning the microwave frequency by ±2.5 MHz (±1.2 MHz), highlighting the noise-resilient feature. These results demonstrate that geometric quantum computation is a potential method for achieving high-fidelity qubit manipulation reproducibly in semiconductor quantum computation.  
\end{abstract}

\maketitle


\section*{\label{sec:level1}Introduction}

Spin qubits based on semiconductor quantum dots \cite{lossQuantumComputationQuantum1998b} are considered among the most promising candidates for building blocks in quantum computers \cite{zhangSemiconductorQuantumComputation2019c,stanoReviewPerformanceMetrics2022}. Among various quantum-dot types, hole spins in germanium combine several features favorable for encoding qubits \cite{scappucciGermaniumQuantumInformation2020}, including all-electrical control based on intrinsic spin-orbit interaction \cite{liuUltrafastElectricallyTunable2023,wangUltrafastCoherentControl2022b,watzingerGermaniumHoleSpin2018,hendrickxFourqubitGermaniumQuantum2021}, low effective mass that simplifies fabrication \cite{lodariLightEffectiveHole2019}, and reduced hyperfine interaction attributed to atomic p orbitals \cite{scappucciGermaniumQuantumInformation2020} and the abundance of net-zero nuclear spin isotopes \cite{hendrickxFourqubitGermaniumQuantum2021,wangUltrafastCoherentControl2022b}. Recently, significant developments have been made, including high-quality single- and two-qubit gates \cite{lawrieSimultaneousSinglequbitDriving2023,wangOperatingSemiconductorQuantum2024a,hendrickxFastTwoqubitLogic2020,hendrickxSweetspotOperationGermanium2024}, singlet-triplet qubits operating at much lower magnetic field \cite{jirovecSinglettripletHoleSpin2021,zhangUniversalControlFour2025a,rooneyGateModulationHole2025} and a 4×4 two-dimensional crossbar quantum dot array with fine tunability \cite{borsoiSharedControl162024}. Despite these achievements, a persistent
issue is that strong spin-orbit interaction is a double-edged sword, enabling ultrafast qubit control while exposing the qubit to complex noise environments \cite{wangOptimalOperationPoints2021,scappucciGermaniumQuantumInformation2020}. In addition, anisotropic susceptibility to noise results in site-dependent qubit properties and nonuniform qubit operations, posing significant obstacles for large-scale quantum computation \cite{hendrickxSweetspotOperationGermanium2024}. As the number of qubits scales up, semiconductor quantum computation is entering the Noisy Intermediate-Scale Quantum (NISQ) era, which demands increasingly stringent requirements for noise-resilient, uniform, and high-fidelity operations across dense arrays \cite{preskillQuantumComputingNISQ2018}. Therefore, it is a paramount challenge to improve error robustness while maintaining high-quality qubit operation \cite{noiriFastUniversalQuantum2022a}.

As a seminal and enduring example, geometric quantum computation (GQC) continues to attract significant attention in the field of quantum computation \cite{pachosNonAbelianBerryConnections1999,duanGeometricManipulationTrapped2001,zuExperimentalRealizationUniversal2014,maSinglespinqubitGeometricGate2024}. Compared to dynamical phases, geometric phases are determined by the geometry of the evolution path and are immune to perturbations that do not change the enclosed area of the path \cite{aharonovPhaseChangeCyclic1987,zhuGeometricQuantumGates2005b}, indicating their robustness against certain types of noise \cite{dechiaraBerryPhaseSpin2003,carolloSpinGeometricPhase2004}. Taking advantage of geometric phases, geometric quantum gates can achieve noise-resilient operations. Early GQC proposals utilized adiabatic evolution to suppress unwanted transitions and state leakage \cite{QuantalPhaseFactors1984,wilczekAppearanceGaugeStructure1984,duanGeometricManipulationTrapped2001,huangExperimentalRealizationRobust2019}. However, coherence times are typically much shorter than the evolution times of adiabatic geometric gates, rendering this approach universally impractical for quantum computation. To speed up gate manipulation while still retaining robustness, non-adiabatic geometric quantum computation (NGQC) has been proposed and demonstrated. Depending on the dimension of the governing Hamiltonian and energy level configurations \cite{duanGeometricManipulationTrapped2001,mousolouUniversalNonadiabaticHolonomic2014b}, geometric gates are classified as Abelian phase-based \cite{zhuImplementationUniversalQuantum2002b,zhuUnconventionalGeometricQuantum2003a,yangExperimentalImplementationShortPath2023} or non-Abelian phase-based \cite{zuExperimentalRealizationUniversal2014,yanExperimentalRealizationNonadiabatic2019}. Notably, Abelian-phase NGQC features experimental simplicity and feasibility, paving a promising way toward universal quantum computation \cite{zhangHighfidelityGeometricGate2020b}. As a two-level system, hole spin qubits are well-suited for implementing NGQC schemes based on Abelian phases, benefiting from ultrafast qubit manipulations \cite{hendrickxFastTwoqubitLogic2020} and easy fabrication of two-dimensional lattice structures \cite{wangOperatingSemiconductorQuantum2024a}. By designing specific evolution paths, NGQC  demonstrates the robustness against off-resonance noise (Larmor frequency noise) and systematic noise (Rabi frequency noise) \cite{guoOptimizingNonadiabaticGeometric2023b,chenErrorTolerantGeometricQuantum2022}. Despite the implementation of Abelian-phase geometric quantum gates, their noise-resilient feature remains unverified and the control fidelities remain far below the threshold for fault-tolerant semiconductor quantum computation \cite{wangExperimentalRealizationNonadiabatic2016,maSinglespinqubitGeometricGate2024}. Nevertheless, achieving high-fidelity, noise-resilient qubit manipulation is essential for the practical application of GQC in scalable quantum computing.

Here, we establish a hole spin qubit based on a strained germanium quantum dot. We achieve a 19 MHz Rabi frequency with a qubit dephasing time of $T_{2}^{*}$ = 136 ns, which can be extended to 6.75 $\upmu$s through dynamical decoupling techniques. To evaluate qubit control performance, we benchmark single-qubit gate control fidelities at a series of Rabi frequencies $f_{\rm{Rabi}}$ utilizing gate set tomography (GST) \cite{nielsenGateSetTomography2021a}. The control fidelities of both the $X$/2 gate and $Y$/2 gate approach approximately 99.9$\%$ as the Rabi frequency increases, while those of $I$ gate saturate at 97.48$\%$, which reveals that off-resonance noise seriously affects qubit control fidelity \cite{mehmandoostDecoherenceInducedSparse2024}. Furthermore, to remove obstacles caused by noise, we introduce non-adiabatic geometric gates based on Abelian geometric phases to implement noise-resilient quantum gates \cite{zhangHighfidelityGeometricGate2020b,xuExperimentalImplementationUniversal2020b}. To clearly compare the performance of GQC with that of dynamical gates, we characterize the control fidelities of geometric gates across a wide range of Rabi frequencies. The geometric gate control fidelities always remain above 99$\%$, demonstrating uniform and high-quality qubit manipulation. Notably, the maximum fidelity can reach 99.9$\%$, satisfying the demand for quantum error correction using surface code \cite{fowlerSurfaceCodesPractical2012}. Moreover, we experimentally demonstrate the noise-resilient feature of geometric quantum gates by detuning the microwave frequency away from the qubit frequency. The control fidelities of geometric $X$/2 and $Y$/2 ($I$) gates can still reach 99$\%$ with the microwave frequency detuned by ±2.5 MHz (±1.2 MHz), highlighting their role in reducing the need for frequent qubit frequency calibration \cite{berrittaPhysicsinformedTrackingQubit2024,zhangHighfidelityGeometricGate2020b}. Our results reveal that GQC is a convincing method for achieving reliable and high-fidelity qubit control in complex noise environments.

\section*{\label{sec:level2}Results}
\begin{figure}[!htbp]
	\includegraphics{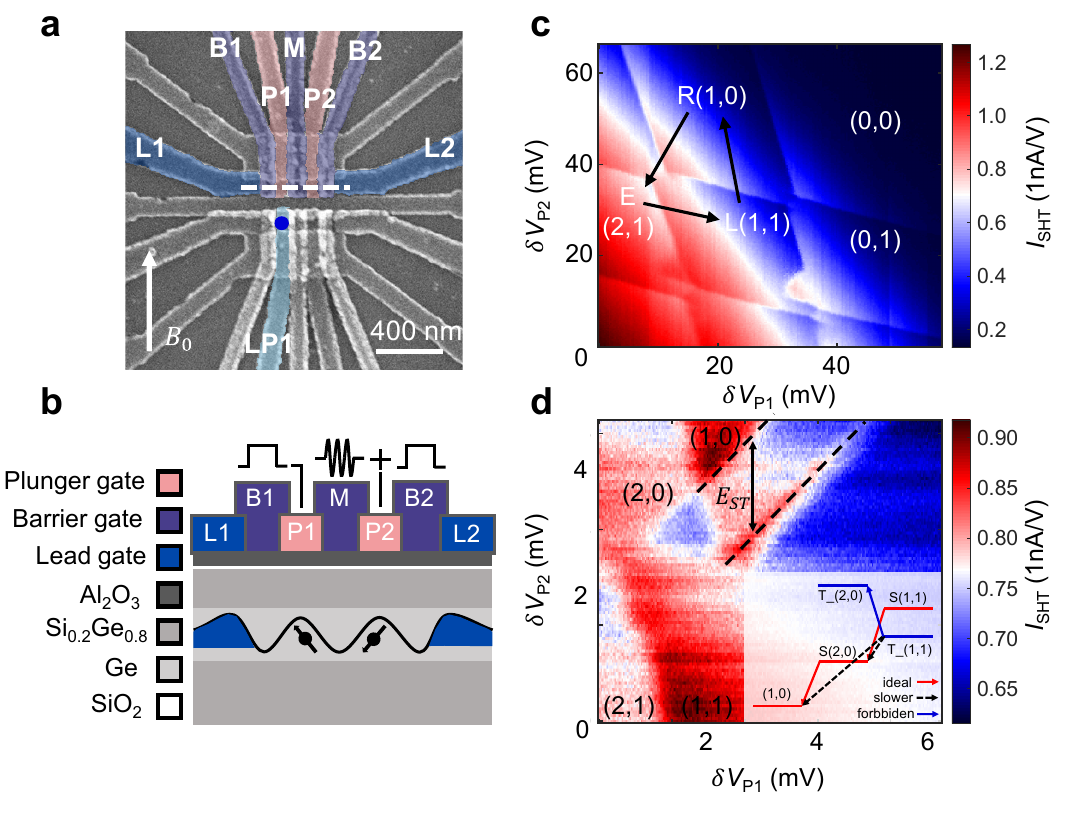}
	\caption{\label{fig:1} \textbf{Device fabrication and experimental setup.} 
	\textbf{a} False-colored SEM images of the device. The upper side is the 
	DQDs structure, and the lower region serves as a charge sensor (blue 
	circle). \textbf{b} Cross-section schematic of the device along the white 
	dashed line in \textbf{a}. Holes confined under gate P1 and P2 can tunnel 
	to the reservoir under gate L1 and L2, with the tunnel rates adjusted by 
	barrier gates B1 and B2. Square pulses and microwave pulses are applied to 
	gates P1 and P2 for state readout and manipulation. \textbf{c} Charge 
	stability diagram of DQDs. (N1, N2) represent the number of holes in the 
	quantum dot under gate P1 and P2. Points E, L and R represent the relative 
	position of square pulses for ELR. \textbf{d} The latching region for ELR 
	is indicated by parallel black dashed lines, where the distance corresponds 
	to $E_{\rm{ST}}$. Inset: energy level and state-ladder schematic at the 
	latching region.}
\end{figure}

\noindent\textbf{Measurement techniques.} Fig.~\ref{fig:1}a presents a false-colored scanning electron microscope (SEM) image of a double quantum dots (DQDs) device fabricated on an undoped strained germanium wafer \cite{kongUndopedStrainedGe2023}. The fabrication details are similar to those described in Ref. \cite{zhangGiantAnisotropySpin2020c}. The device consists of two sections: the upper part is configured as the DQDs, and the lower portion (blue circle) serves as a charge sensor to detect the charge configuration of DQDs. An external in-plane magnetic field $B_{0}$ of 1333 mT is applied to provide Zeeman splitting for defining spin qubits. Fig.~\ref{fig:1}b shows a cross-section schematic of the DQDs along the white dashed line in Fig.~\ref{fig:1}a. The hole spins are confined in DQDs by selectively tuning the plunger gates P1 and P2, as well as barrier gates. Additionally, the tunnel coupling between the quantum dots and their reservoirs, and inter-dot tunnel coupling can be modified separately via three barrier gates B1, B2 and M. We apply microwave pulses to gate P2, and three-step pulses to gates P1 and P2 to initialize, control and read out the qubit state (see Methods for details). Detected by the charge sensor, the charge stability diagram of DQDs is depicted in Fig.~\ref{fig:1}c, where Ni represents the number of holes in the quantum dot under gate Pi, respectively.

We select the (1, 1) - (2, 0) inter-dot tunneling region for spin state readout via enhanced latching readout (ELR) \cite{harvey-collardHighFidelitySingleShotReadout2018a} which is based on Pauli spin blockade (PSB) \cite{johnsonTripletSingletSpin2005a}. The latching phenomenon originates from the asymmetric coupling strengths between the two dots and their adjacent reservoirs \cite{harvey-collardHighFidelitySingleShotReadout2018a}. By tuning the voltage on gate B2, the tunnel rate between dot beneath P2 and the right reservoir beneath L2 can be pinched off (Supplementary Note 1), forcing holes in dot beneath P2 to tunnel into the left reservoir via co-tunneling with dot beneath P1, a process that also occurs slowly.

The ELR process is explained by comparing the theoretical latching behaviors 
when the spin state is initialized as a singlet or triplet state 
\cite{maSinglettripletstateReadoutSilicon2024}. The energy levels and 
state-ladder schematic in the latching region are shown in the inset of 
Fig.~\ref{fig:1}d, where only the singlet state $S$ and the ground triplet 
state $T_{-}$ are considered for simplification. If $T_{-}$(1, 1) is 
initialized, it cannot tunnel into $S$(2, 0) due to PSB, or into $T_{-}$(2, 0) 
due to insufficient energy. Direct tunneling into (1, 0) is suppressed due to a 
slow tunnel rate, resulting in no detectable signal in the latching region. In 
contrast, the singlet state tunnels to (1, 0) via (2, 0) state, thereby 
generating a full-hole signal. Therefore, the singlet and triplet states can be 
clearly distinguished by monitoring hole tunneling from dot beneath P1 to the 
left reservoir. The following experiments employ single-shot ELR. 
Fig.~\ref{fig:1}c illustrates the cyclic E-L-R pulse sequence used to probe the 
latching region. The spin state is initialized into a S-T mixed state at stage 
L via the transition from (2, 1) to (1, 1). Scanning the voltage of stage R 
within the range shown in Fig.~\ref{fig:1}d, the latching region in (1, 0) 
becomes visible, as depicted by parallel black dashed lines. The energy 
difference between the singlet and triplet states is 0.8 meV, determined by a 
lever arm of 0.13 eV V$^{-1}$ (see Supplementary Note 2).

\vspace{1em} 

\begin{figure}[!htbp]
	\includegraphics{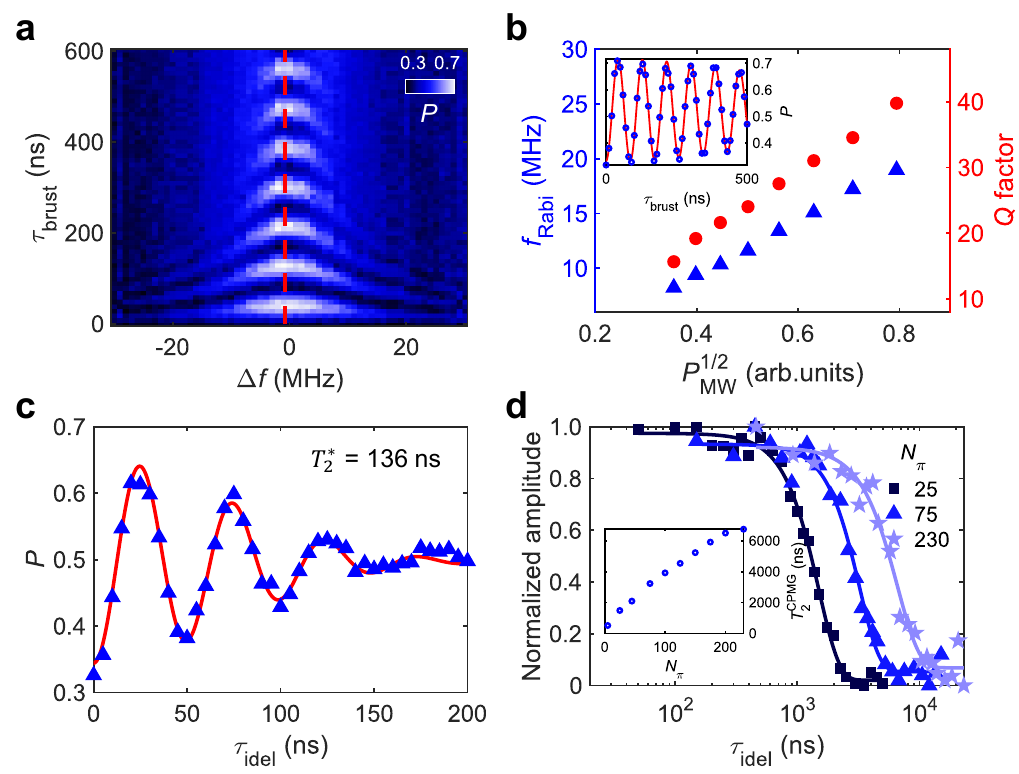}
	\caption{\label{fig:2} \textbf{Qubit properties.} \textbf{a} The Rabi 
	chevron pattern: singlet state probability $P_{S}$ as a function of $\tau 
	_{\rm burst}$ and $\Delta f$. \textbf{b} Rabi frequency (blue triangles) 
	and the Q factor (red circles) measured under different microwave power P, 
	showing no saturation. The largest $f_{\rm Rabi}$ of 19 MHz and $Q$ factor 
	of 39.78 are achieved. Inset: $P_{S}$ as a function of $\tau _{\rm burst}$ 
	along the red dashed line in \textbf{a}. The red solid line represents the 
	fitting result, yielding a Rabi frequency of 11.56 MHz and $Q$ factor of 
	18.04. \textbf{c} Ramsey fringe obtained by varying the idle time $\tau 
	_{\rm{idle}}$, with the fitted dephasing time $T_{2}^{*}$ of 136 ns. The 
	red solid line represents the fitting result. \textbf{d} Coherence decay 
	under CPMG sequences as a function of total idle time for different 
	$N_{\pi}$. Inset: $T_{2}^{\rm CPMG}$ as a function of $N_{\pi}$. For the 
	largest $N_{\pi}=230$, we obtain $T_{2}^{\rm CPMG}$ = 6.75 $\upmu$s.}
\end{figure}

\noindent\textbf{Qubit Properties.} Next, we focus on coherent qubit control. 
The spin state is initialized into $T_{-}$ at stage L by waiting for a 
sufficiently long period (see Supplementary Note 3). Qubit operations are 
performed via electric dipole spin resonance (EDSR) mediated by the intrinsic 
spin-orbit interaction (SOI) 
\cite{wangUltrafastCoherentControl2022b,hendrickxFastTwoqubitLogic2020}. When 
the microwave frequency ($f_{\rm MW}$) matches the qubit resonance frequency, 
$T_{-}$ is operated to $S$ and PSB is lifted. Consequently, a full-hole signal 
can be detected at stage R 
\cite{maSinglettripletstateReadoutSilicon2024,harvey-collardHighFidelitySingleShotReadout2018a}.
 By applying a rectangular-shaped microwave burst ($\tau_{\rm{burst}}$) with a 
fixed duration of $t_{\uppi}$ (time for a ${\pi}$ rotation), two resonance 
frequencies $f_{1}$ = 5.60 GHz and $f_{2}$ = 7.53 GHz are measured 
(Supplementary Note 4). Using the relation $hf_{i} = g_{i}\mu_{B}B_{0}$, 
g-factors $g_{1}$ = 0.29 and $g_{2}$ = 0.39 are extracted, where $h$ is 
Planck’s constant and $\mu_{B}$ is the Bohr magneton. In the following 
measurements, we mainly focus on the qubit with the higher frequency and 
characterize its properties \cite{hendrickxFourqubitGermaniumQuantum2021}. 
Firstly, Fig.~\ref{fig:2}a shows the Rabi chevron pattern by varying 
$\tau_{\rm{burst}}$ and $\Delta{f}$ (defined as $f_{\rm{MW}}-f_{2}$). Along the 
red dashed line where $f_{\rm{MW}}$ resonates with $f_{2}$ (inset of 
Fig.~\ref{fig:2}b), the hole spin rotates over an angle determined by 
$\tau_{\rm{burst}}$. Fitting the oscillation yields Rabi frequency $f_{\rm 
Rabi}$ = 11.56 MHz, $T_{2}^{\rm Rabi}$ = 1.56 $\upmu$s, and a quality factor 
$Q$ = 18.04 (defined by $2*T_2^{\rm{Rabi}}*f_{\rm{Rabi}}$). As shown in 
Fig.~\ref{fig:2}b, increasing driving power continuously, both $f_{\rm Rabi}$ 
and the $Q$ factor rise monotonically without saturation, as indicated by the 
bule triangles and red circles, respectively. At maximum driving power, $f_{\rm 
Rabi}$ reaches 19 MHz with the highest $Q$ = 39.78. To characterize the qubit 
dephasing time $T_{2}^{*}$, we perform a Ramsey experiment with $\Delta{f}$ = 
20 MHz. By varying the idle time ($\tau_{\rm idle}$) between two $X$/2 pulse, 
the Ramsey decay (see Fig.~\ref{fig:2}c) yields a dephasing time $T_{2}^{*} = 
136$ ns at $f_{\rm Rabi} = 11.56$ MHz. $T_{2}^{*}$ remains nearly constant 
across the measured $f_{\rm Rabi}$ range (Supplementary Note 5). Furthermore, 
we apply the Carr–Purcell–Meiboom–Gil (CPMG) sequence to alleviate the effect 
of noise and extend the coherence time \cite{cywinskiHowEnhanceDephasing2008b}, 
in which a series of $Y$ gates are inserted between two $X$/2 operations to 
refocus the qubit state \cite{veldhorstAddressableQuantumDot2014c}. 
Fig.~\ref{fig:2}d shows a set of normalized signals with a coherence time 
$T_{2}^{\rm CPMG}$ of 6.75 $\upmu$s for ${N}_{\pi} = 230$ (${N}_{\pi}$ 
denotes the number of $Y$ gates in the CPMG sequence), which is 50 times longer 
than $T_{2}^{*}$. As shown in the inset of Fig.~\ref{fig:2}d, $T_{2}^{\rm 
CPMG}$ increases linearly with ${N}_{\pi}$, which indicates that low-frequency 
noise is one of the important noise sources, as revealed through the equivalent 
filter functions of CPMG pulse sequence 
\cite{cywinskiHowEnhanceDephasing2008b,hendrickxSweetspotOperationGermanium2024,madzikControllableFreezingNuclear2020b}.

\vspace{1em} 

\noindent\textbf{Control fidelity.} Among the various quantum benchmarking 
techniques, randomized benchmarking (RB) 
\cite{knillRandomizedBenchmarkingQuantum2008b} and gate set tomography (GST) 
\cite{blume-kohoutDemonstrationQubitOperations2017,nielsenGateSetTomography2021a}
 stand out for their robustness against state preparation and measurement 
(SPAM) errors. We therefore implement both techniques to accurately assess the 
qubit control fidelity. RB experiment employs two types of sequences: reference 
sequences and interleaved sequences. The reference sequence consists of $N$ 
random Clifford gates and an additional recovery gate that recovers the spin to 
the spin-down state. The interleaved sequences, created by inserting the target 
gate between adjacent random gates of the reference sequences, are used to 
evaluate the control fidelity of the target gate (More details shown in 
Supplementary Note 6). Fig.~\ref{fig:3}a shows the control fidelities of 
single-qubit gates measured using RB. The average single-qubit gate fidelity is 
99.49$\%$, with both $I$, $X$/2 and $Y$/2 gate control fidelities exceeding 
99$\%$ - surpassing the threshold for quantum error correction using surface 
codes \cite{fowlerSurfaceCodesPractical2012}. Here, the $I$ gate denotes an 
identity operation implemented by idling for the duration of a $\pi$/2 $X$ or 
$Y$ rotation. For GST, the default gate set contains $I$, $X$/2 and $Y$/2 
gates. GST sequences consist of two components: fiducials for state preparation 
and germs for amplifying various types of gate errors. The number of total 
sequences depends on sequence depth. We select a sequence depth of 8, yielding 
a total of 448 GST sequences. Each sequence is constructed by interleaving 
these fiducials with germ sequences, after which we measure the singlet state 
probability (More details shown in Supplementary Note 6). Analyzed by the 
python package pyGSTi, Fig.~\ref{fig:3}b presents the single-qubit gate 
fidelities across different $f_{\rm Rabi}$ characterized by GST 
\cite{nielsenProbingQuantumProcessor2020}. Gate fidelities show a positive 
correlation with $f_{\rm Rabi}$, in agreement with the $Q$ factor trend shown 
in Fig.~\ref{fig:2}b. The best performance occurs at $f_{\rm Rabi}$ = 19 MHz, 
achieving control fidelity of 97.48$\%$ for $I$ gate, 99.81$\%$ for $X$/2 gate 
and 99.88$\%$ for $Y$/2 gate. Note that the $I$ gate fidelities remain 
substantially below 99$\%$ across the entire $f_{\rm Rabi}$ range, which is 
mainly due to the shorter $T_2^*$ compared to the idle time composed of 
multiple consecutive $I$ gates in the GST sequence. On the one hand, the 
infidelity owing to Rabi decay is small, as indicated by the high fidelity 
(99.49$\%$ on average) single-qubit gate characterized by RB 
\cite{noiriFastUniversalQuantum2022a}. On the other hand, some pulse sequences 
in the GST experiment include idle times much longer than $T_{2}^{*}$, which 
mainly limits the control fidelity \cite{xueQuantumLogicSpin2022a}. Raising 
$f_{\rm Rabi}$ directly helps suppress the effect of dephasing 
\cite{noiriFastUniversalQuantum2022a,wangOperatingSemiconductorQuantum2024a}. 
As a result, the control fidelity of $I$ gate improves 3.46$\%$ and those of 
$X$/2 and $Y$/2 gate are much closer to 99.9$\%$. Unfortunately, when $f_{\rm 
Rabi}$ exceeds 15 MHz, the $I$ gate fidelity appears to saturate but still 
significantly below 99$\%$. TABLE \ref{tab:t1} lists the control fidelities of 
dynamical gates characterized by RB and GST at $f_{\rm Rabi} = 19$ MHz, which 
are similar to each other except for the $I$ gate. This discrepancy arises from 
the differences in sequence composition between the two methods (see 
Supplementary Note 6) \cite{wangOperatingSemiconductorQuantum2024a}.


\vspace{1em} 

\noindent\textbf{Geometric quantum computation.} Analysis of the error generator in GST reveals that the errors can be classified into coherent Hamiltonian errors and incoherent stochastic errors \cite{huangPerformanceQuantumError2019,blume-kohoutTaxonomySmallMarkovian2022}. For coherent errors, diverse strategies can be employed to mitigate them, such as dynamical decoupling or recalibrating control \cite{huangPerformanceQuantumError2019,dehollainOptimizationSolidstateElectron2016,blume-kohoutTaxonomySmallMarkovian2022}. For stochastic errors, such as dephasing and depolarizing noise, we propose a noise-resilient operation strategy to counteract uncorrelated noise \cite{huangPerformanceQuantumError2019}. We introduce GQC, a technique known for its superior robustness \cite{chenUniversalRobustGeometric2024,zhangHighfidelityGeometricGate2020b} to resist various noises, enabling uniform and high-fidelity qubit manipulation. 

We consider a Hamiltonian for a two-level qubit system driven by a classical microwave field,

\begin{equation}
H = \frac{h}{2}
\left(
\begin{array}{cc}
\Delta f & (1+\delta) f_{\rm{Rabi}}e^{-i\Phi } \\
(1+\delta) f_{\rm{Rabi}}e^{i\Phi } & -\Delta f \\
\end{array}
\right),
\end{equation}

where $\Delta f$  represents the difference between $f_{\rm{MW}}$ and $f_{2}$. The corresponding terms appear in the diagonal elements of the Hamiltonian. The fluctuations in $f_{\rm Rabi}$, $\delta $, manifest in the off-diagonal elements of the Hamiltonian.

To construct nonadiabatic single-qubit geometric gates, the evolution loop is designed as a closed trajectory consisting of three intervals in the experiment: $0 {\rightarrow} \tau_{1}$,$\tau _{1}{\rightarrow} \tau _{2}$,$\tau _{2}{\rightarrow} \tau _{3}$. During each interval, the microwave has different phases and amplitudes, as detailed below. 

\begin{equation}
\begin{aligned}
&\int_{0}^{\tau _{1}}2\pi f_{\rm{Rabi}} dt=\theta ,\Phi (t)= \phi -\pi /2, t\in [0,\tau _{1}], \\
&\int_{\tau _{1}}^{\tau _{2}}2\pi f_{\rm{Rabi}} dt=\pi ,\Phi (t)= \phi +\gamma -\pi/2, t\in [\tau _{1},\tau _{2}], \\
&\int_{\tau _{2}}^{\tau _{3}}2\pi f_{\rm{Rabi}} dt=\pi -\theta ,\Phi (t)= \phi -\pi /2, t\in [\tau _{2},\tau _{3}].
\end{aligned}
\end{equation}
The formulas above describe one of the proposals to realize a geometric gate, labeled as “Path2” in the following (see the evolution path shown in Fig.~\ref{fig:3}c). An alternative proposal, labeled as “Path1”, replaces the phase of the second interval with $\Phi (t)= \phi +\gamma +\pi/2$. These two proposals target different types of noise. Path1 addresses systematic noise $\delta $, which represents fluctuations of $f_{\rm Rabi}$, while Path2 resists off-resonance noise $\Delta f$, defined by the fluctuations of $f_{2}$ \cite{zhangHighfidelityGeometricGate2020b}.

Following the cyclic evolution loop above, the equivalent evolution operator is 

\begin{equation}
\begin{split}
    U (\theta,\gamma,\phi)&= \cos\gamma +i \sin\gamma  
        \left(
    \begin{array}{cc}
      \cos\theta  & \sin\theta e^{-i\phi} \\
       \sin\theta e^{i\phi}  & -\cos\theta 
    \end{array} \right) \\
    &= \mathrm{exp}(i\gamma\mathbf{n}\cdot\boldsymbol{\sigma}),  
\end{split}
\end{equation}

which determines the rotation around the axis $\textbf{n}=(\sin\theta\cos\phi,\sin\theta\sin\phi,\cos\theta)$ by an angle $-2\gamma$, where $\boldsymbol{\sigma}$ represents the Pauli operators. An arbitrary single-qubit geometric gate can be achieved by adjusting the parameters in the evolution loop. For example, by setting $\theta=\pi/2,\phi=\pi /2,\gamma=0$, a geometric $I$ gate is obtained. This geometric $I$ gate effectively implements a $2\pi$ rotation about x axis through three steps.

Fig.~\ref{fig:3}d presents the control fidelities of single-qubit geometric gates Path2 as a function of $f_{\rm Rabi}$, analyzed through GST. In contrast to dynamical gates, the average fidelities of geometric gates exceed 99$\%$ throughout the entire measurement range, highlighting the superiority of GQC. The maximum control fidelities reach 99.98$\%$, 99.80$\%$, and 99.97$\%$ for the $I$, $X$/2, and $Y$/2 gates, respectively. The detailed data shown in Fig.~\ref{fig:3}d are provided in Supplementary TABLE I. Especially, the fidelity of the $I$ gate can be significantly improved even when $f_{\rm Rabi}$ is small. Unlike Path2, the control fidelity of the gate set achieved by Path1 is well below 99$\%$ (Supplementary Note 11). This is probably because the systematic noise targeted by Path1 is relatively weaker than the off-resonance noise here (Supplementary Note 8) \cite{kawakamiGateFidelityCoherence2016b,yangSiliconQubitFidelities2019b,maSinglespinqubitGeometricGate2024}. To simultaneously resist both types of noise, we introduce an enhanced geometric gate featuring a 3$\pi$ evolution path (named Path3$\pi$) that extends 1.5 times longer than Path1 and Path2 \cite{guoOptimizingNonadiabaticGeometric2023b}. However, its performance is significantly worse than that of Path2, probably due to the longer operation time (see Supplementary Note 9).

\vspace{1em} 

\begin{figure*}[htbp]
	\includegraphics{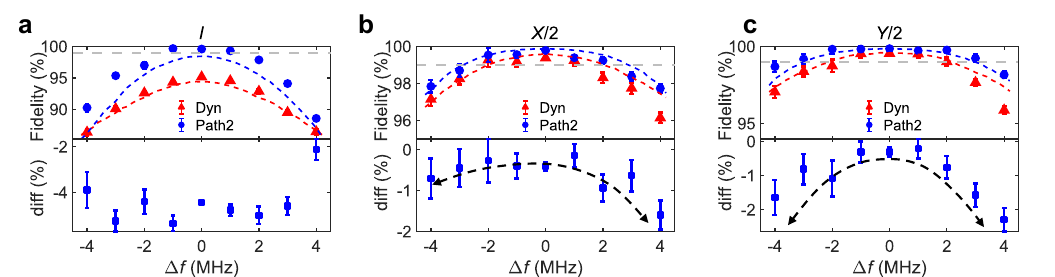}
	\caption{\label{fig:4} \textbf{Noise-resilient feature of single-qubit 
	geometric gates.} The upper panels of \textbf{a-c} show that control 
	fidelities of the $I$, $X$/2 and $Y$/2 gates, realized by both geometric 
	and dynamical approaches, as a function of $\Delta f$. The light grey lines 
	represent 99$\%$ control fidelity threshold. Notably, even with detuning 
	$\Delta f$ by ±2.5 MHz (±1.2 MHz), the control fidelities of geometric 
	$X$/2 and $Y$/2 ($I$) gates can still reach 99$\%$. The differences in the 
	control fidelities between the two approaches are displayed in the lower 
	panels. The black dashed arrows serve as visual guides to emphasize that as 
	$\Delta f$ increases, the control fidelities of the dynamical gates 
	decrease faster than those of the geometric gate. Error bars represent the 
	95$\%$ confidence level. The experimental results are consistent with the 
	numerical simulations (blue and red dashed lines).}
\end{figure*}

\noindent\textbf{Noise-resilient operation.} After realizing high-fidelity single-qubit geometric gates, we further demonstrate their noise-resilient characteristics against $\Delta f$. As shown in the upper panels of Fig.~\ref{fig:4}, we compare the performance of the $\Delta f$- error-affected geometric gate (red triangles) with that of the corresponding dynamical counterparts (“Dyn” for short, blue circles). As expected, the gate set implemented with geometric gate Path2 is clearly superior to the conventional dynamical gate. The fidelities of the geometric $X$/2 and $Y$/2 ($I$) gates exceed 99$\%$ even when $\Delta f$ = ±2.5 MHz (±1.2 MHz). Furthermore, $\Delta f$ measurements over 55 hours fall precisely within this range, verifying the ability of the geometric gate to alleviate the need for frequent calibration of qubit resonance frequency (Supplementary Note 10). Moreover, the lower panels of Fig.~\ref{fig:4} show the differences in control fidelities between the geometric and dynamical gates. As $|\Delta f|$ increases, the fidelities of the dynamical gate decrease faster than those of the geometric gate, which confirms that the geometric gate is more robust against off-resonant noise. Additionally, the robustness against Rabi error $\delta $ is displayed by comparing Path1 with the dynamical gate in Supplementary Note 11. In order to explain the noise-resilient feature of the geometric gate, we numerically simulate the fidelity with additional off-resonance noise. Based on the $T_2^*$ values and the GST report, we choose an appropriate noise strength as the starting point. We then calculate the probability of the singlet state for each GST sequence, and analyze the fidelity. By numerical simulation, the final parameters we extracted are $\delta f^{\rm{Larmor}}$ = 1.80 MHz and $\delta f^{\rm{Rabi}}$ = 58.48 kHz. The calculated fidelities, overlaid in Fig.~\ref{fig:4} (Supplementary Note 7), show that the theoretical results align with the experimental results in terms of overall trends. These results clearly demonstrate the advantages of GQC in achieving high-fidelity qubit manipulations despite the presence of complex noise.

\section*{\label{sec:level6} Discussion}

For the practical implementation of geometric quantum gates, two necessary considerations are optimal scheme selection and geometric gate operation time. The geometric gates we have implemented are tailored for different noise types. Therefore, the best scheme choice depends on the specific noise experienced by spin qubits. In semiconductor quantum computation, the dominant noise source depends on the wafer material, device architecture, and frequency domain of interest \cite{burkardSemiconductorSpinQubits2023}. Through characterization of qubit coherence time, noise spectrum analysis \cite{yonedaQuantumdotSpinQubit2018c,dumoulinstuyckSiliconSpinQubit2024}, report analyzed by gate set tomography \cite{blume-kohoutTaxonomySmallMarkovian2022}or the envelope of Rabi oscillation decay \cite{PhysRevX.10.011060}, we can make a basic assessment of noise type and noise level, enabling the efficient selection of the optimal geometric gate scheme. Otherwise, counterproductive effects may arise (Supplementary Note 8). Depending on the relative strengths of systematic and off-resonance noise, we can choose either Path1 or Path2. When the strengths of off-resonance noise and systematic noise are comparable, alternative protocols such as Path3$\pi$ or nonadiabatic geometric quantum computation with noncyclic evolution paths can be employed \cite{maNoncyclicNonadiabaticGeometric2023}. On the other hand, the evolution times of all types of geometric gates implemented here are four or six times longer than those of dynamical gates. There is a trade-off between extended operation time and enhanced noise resistance. Encouragingly, a noncyclic nonadiabatic geometric quantum gate was recently proposed to shorten the evolution time, effectively avoiding the cumulative disturbances of errors due to excessive time consumption \cite{maNoncyclicNonadiabaticGeometric2023,yangExperimentalImplementationShortPath2023}.

Regarding the practical advantages of geometric gate applications, reproducible high-fidelity qubit control without frequent recalibration of control parameters is particularly attractive. In current devices with a limited number of qubits, calibration requires a considerable amount of time \cite{noiriFastUniversalQuantum2022a,philipsUniversalControlSixqubit2022c,dumoulinstuyckSiliconSpinQubit2024}. Although improvements in circuit optimization and fabrication techniques can help mitigate the noise strengths, complete elimination of noise remains challenging. Expanding to large-scale quantum computation, the varying noise environments experienced by individual qubits pose significant challenges to reproducibly achieve high-fidelity qubit control. As calibration becomes increasingly resource-demanding, it significantly hinders the efficiency of the qubit processors. Geometric gates can potentially reduce both the time and resource consumption required for recalibration in large-scale quantum computation.

In conclusion, we characterize the single-hole spin qubit in germanium DQDs and extract its control fidelity via GST. Along with increasing $f_{\rm Rabi}$ and the $Q$ factor, the fidelities of the dynamical gate gradually saturated and those of $X$/2 and $Y$/2 approach approximately 99.9$\%$. However, for the $I$ gate, the fidelity remains significantly below 99$\%$ which indicates the susceptibility of spin qubit to off-resonance noise. In order to explore a strategy for achieving noise-resilient and high-fidelity spin qubit manipulation in future large-scale quantum computation, we introduce GQC featuring noise robustness. Two different geometric evolution loops, named as Path1 and Path2, are implemented to specifically deal with systematic and off-resonance noise, respectively. Regarding geometric gate Path2, the average control fidelity of the gate set consistently reach 99$\%$ across a wide range of $f_{\rm Rabi}$, demonstrating its potential to achieve uniform qubit control under various situations. In addition, the $I$, $X$/2, and $Y$/2 gates exhibit maximum control fidelities of 99.98$\%$, 99.80$\%$, and 99.97$\%$, respectively, despite requiring four times the operation time of dynamical gates. Furthermore, we demonstrate the robustness of geometric gates by adding off-resonance noise $\Delta f$, which currently has a more harmful impact on qubit properties. The error-affected performance of Path2 is superior to that of dynamical gates throughout the entire measurement. Calibration of the qubit resonance frequency in experiments is a time-intensive and resource-demanding process. The geometric gate, despite detuning $\Delta f$ by ±2.5 MHz (±1.2 MHz), still achieves fidelities of the $X$/2 and $Y$/2 ($I$) gates that reach fault-tolerant quantum computing thresholds, confirming its ability to preform high-fidelity and noise-resilient qubit operations while alleviating the need for frequent qubit frequency calibration. Consequently, the demonstrated geometric quantum gate has broad application prospects for implementing high-fidelity and robust qubit manipulations for large-scale quantum computation. Moreover, in future work, combined with two-qubit geometric gates \cite{zhangHighfidelityGeometricGate2020b, PhysRevA.111.042609}, it is expected to realize robust universal geometric quantum computation in a large-scale qubit array.

\section*{Methods}
We perform the experiments using an Oxford dry dilution refrigerator with a base temperature of approximately 15 mK. A three-stage gate voltage pulse is generated by an arbitrary waveform generator (AWG Keysight M8190). The pulse is combined with the DC-voltages using an analog summing amplifier (SRS SIM980) and applied to plunger gates, which are connected to the DC-port of a commercial bias tee (Anritsu K251). Meanwhile, leveraging I/Q modulated signals from Tektronix AWG5208 channel pairs, we apply a microwave signal generated by a vector signal generator (Keysight E8267D) connected to the Rf-in port of the bias tee. The microwave pulse modulation, generated by the AWG5208, is turned on 500 ns before and turned off 500 ns after the microwave signal. By measuring the current of the single-hole transistor, we can detect the qubit state by a charge sensing technique. Being Amplified by room-temperature amplifier (SRS SR570 and SR560), the output current can be digitized by a PCI-based waveform digitizer (AlazarTech ATS 9440) at a sampling rate of 5 MSa/s.

\section*{Data availability}
The data used in this study are available in the Zendo database under accession code 10.5281/zenodo.16450274. Source Data are provided with this paper.

\section*{Code availability}
Error taxonomy was performed with the pyGSTi package \cite{nielsenProbingQuantumProcessor2020,blume-kohoutTaxonomySmallMarkovian2022}. All other supporting algorithms
are provided in the main text and Supplementary Material through equations.

\section*{References}
\bibliography{apssamp}

\section*{Acknowledgments}
This work was supported by the National Natural Science Foundation of China (Grants Nos.92165207, 12474490, 12034018, 92265113 and 62404248), the Innovation Program for Quantum Science and Technology (Grant No. 2021ZD0302300). This work was partially carried out at the USTC Center for Micro and Nanoscale Research and Fabrication.

\section*{Author contributions}
Y.-C.Z. performed the bulk of measurement and data analysis with the help of R.-L.M. Y.-C.Z. fabricated the device with the help of Y.L., H.-T.J. and Z.-T.W. Z.-Z.K. and G.-L.W.
supplied the planar germanium heterostructure. Y.-C.Z. wrote the manuscript with inputs from other authors. A.-R.L. provided theoretical support and contributed to the simulation. H.-O.L., C.-X.Z. and G.-C.G. advised on experiments. X.Z. and G.C. helped prepare the manuscript. H.-O.L. and G.-P.G. supervised the project.

\section*{Competing Interests}
The authors declare no competing interests.

\section*{Tables}

\begin{table}[!h]
	\renewcommand{\arraystretch}{1.5}
	\setlength{\tabcolsep}{5pt}
	\centering
	\caption{The fidelity of the dynamical single$\-$qubit gate set} 
	\label{tab:t1}
	
	\begin{tabular}{|c|c|c|c|c|}
		\hline 
		& Ref. & $I$ & $X$/2 & $Y$/2\\ 
		\hline 
		RB & 99.34(8)$\%$ & 99.21(3)$\%$ & 99.63(2)$\%$ & 99.62(2)$\%$ \\
		\hline
		GST & -- & 97.48(9)$\%$ & 99.81(9)$\%$ & 99.88(8)$\%$ \\
		\hline 
	\end{tabular}
	
\end{table}

\section*{Figure Legends/Captions}

\begin{figure}[!htbp]
	\includegraphics{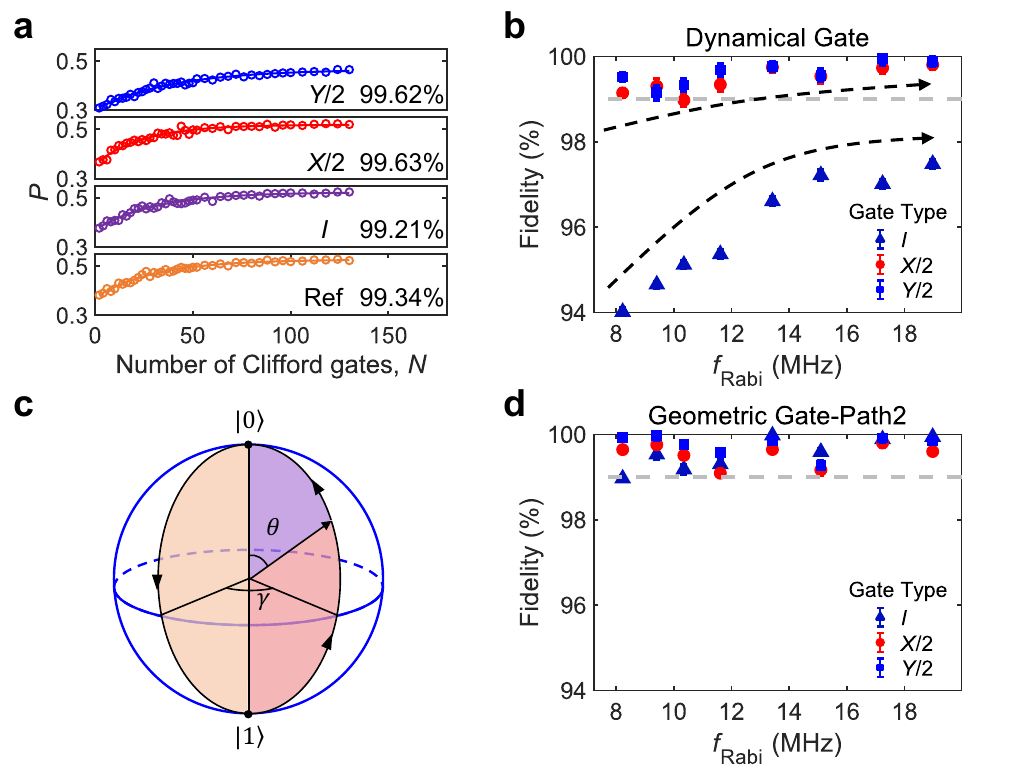}
	\caption{\label{fig:3} \textbf{Control fidelity and geometric quantum 
	computation.} \textbf{a} Sequence fidelities for reference (abbreviated as 
	“Ref”) and each interleaved characterized by RB, where N represent the 
	number of Clifford gates in the reference sequences. The control fidelity 
	is shown below the corresponding curves. \textbf{b} The control fidelities 
	of the dynamical gate set extracted by GST as a function of $f_{\rm Rabi}$. 
	The black dashed arrows guide the eyes, indicating the tendency of the 
	fidelity. The light grey line represents the 99$\%$ control fidelity 
	threshold. As the Rabi frequency increases, the control fidelities of $X/2$ 
	and $Y$/2 gates gradually improve to nearly 99.9$\%$, whereas the $I$ gate 
	control fidelity saturates at 97.48$\%$. Error bars represent the 95$\%$ 
	confidence level. \textbf{c} The schematic of the evolution path Path2. 
	\textbf{d} The control fidelities of the geometric gate Path2 as a function 
	of $f_{\rm Rabi}$. The average control fidelity exceeds 99$\%$ at each 
	$f_{\rm Rabi}$, demonstrating the superiority of geometric gates. The light 
	grey line denotes the 99$\%$ control fidelity threshold, consistent with 
	panel \textbf{b}. The data points consistently exceed the reference line at 
	each $f_{\rm Rabi}$. This demonstrates the superior performance of the 
	geometric gates. Error bars represent the 95$\%$ confidence level.} 
\end{figure}

\end{document}